\documentstyle[12pt]{article}
\def\beq{\begin{equation}}   \def\eeq{\end{equation}}
\begin{document}
\begin{titlepage}

\begin{flushright}
TPI-MINN-99/02-T\\
UMN-TH-1738/99\\
NYU-TH-99/1/01\\
hep-th/9901111\\
\end{flushright}

\vspace{0.3cm}

\begin{center}
\baselineskip25pt

{\Large\bf 
Surviving on the Slope:
Supersymmetric Vacuum in the Theories Where It Is Not Supposed 
to Be}

\end{center}

\vspace{0.3cm}

\begin{center}
\baselineskip12pt

{\large G. Dvali}

\vspace{0.2cm}
Physics Department, New York University, New York, NY 10003

{\em and}

International Center for Theoretical Physics, Trieste, I-34014, Italy
\vspace{0.2cm}

{\em and}

\vspace{0.3cm}
{\large  M.~Shifman} 

\vspace{0.2cm}
Theoretical Physics Institute, University of Minnesota, Minneapolis, 
MN 55455

\vspace{2.5cm}

{\large\bf Abstract} 

\vspace*{.25cm}

\end{center}

In supersymmetric models with the  run-away vacua or with the stable
 but 
{\em non}-supersymmetric ground state there exist stable field
configurations (vacua) which  restore one half of supersymmetry and
are characterized by  constant positive energy density. The formal
foundation for such vacua  is provided by the central extension of the
${\cal N}=1$ superalgebra with the infinite central charge.

\end{titlepage}

In Ref.~\cite{DS1} we found a class of unconventional solutions,  
which exist in supersymmetric theories with a vacuum moduli space,
and are characterized by (i) constant energy density; (ii) topological 
stability. They can be considered as a limiting case of 
the domain walls
(sometimes  we deal with the so-called constant phase 
configurations, sometimes with the winding phase configurations, see 
Sec. 
5 of~\cite{DS1}). One half of supersymmetry may or may not be 
preserved on these solutions. Because of their topological stability 
they can become  vacua of a theory  breaking a part of the Lorentz 
invariance and supersymmetry.  Thus, this is a particular realization 
of the dynamical compactification, the idea central in Ref.~\cite{DS1}.

The existence of the topologically stable vacua with the purely 
gradient (constant) energy density is intuitively clear in the 
examples considered in Ref.~\cite{DS1} since in these examples there 
exist moduli forming a continuous manifold of  supersymmetric 
vacuum states. Being physically inequivalent, the vacua  are 
degenerate, the vacuum energy density vanishes. The vacuum 
moduli spaces occur frequently in supersymmetric theories.

Here we discuss another class of supersymmetric theories,
which, within the standard understanding, have {\em no} 
supersymmetric vacua at all: either there is no vacuum whatsoever 
(the scalar potential has a ``run-away" behavior), or the vacuum state 
exists
but is {\it non-supersymmetric}.
Our solutions  restore a part of  supersymmetry.
This pattern 
is quite  unusual -- normally, stable field configurations with
higher energy density have less supersymmetry than the ground state.
Our analysis shows that the opposite situation is also possible.

 First, we will consider the  so-called 
run-away theories.
The most well-known example of this type is SU(2) SQCD with one 
massless flavor~\cite{ADS} (in general,
SU($N$) SQCD with $N-1$ flavors). At any finite values 
of fields the minimal  
energy is not achieved. One approaches the vanishing energy density 
at infinitely distant points in the space of fields.
So, there is no vacuum in the
conventional sense of this word.  Then, we will 
consider models with the spontaneous breaking of supersymmetry of 
the O'Raifeartaigh type~\cite{OR}.  Both phenomena
 are quite common in the zoo of  
supersymmetric theories. The O'Raifeartaigh models appear
as a low-energy limit of various gauge models producing
the dynamical supersymmetry breaking (for a recent review 
see~\cite{review}).

 We will show that in both cases
BPS saturated solutions exist; they are stable under all localized 
perturbations, preserve one half of the original supersymmetry and, 
thus, present supersymmetric vacuum states. It may well happen 
that such solutions in the future will become a component of a 
phenomenologically successful scenario (e.g.~\cite{DSnew}). 

Let us start from the run-away vacua. Many models with the  
run-away vacuum were considered in the literature. For definitness 
we focus on models with the logarithmic superpotential for the 
moduli,
\beq
{\cal L} = \frac{1}{4} \int d^2 \theta d^2 \bar\theta
\bar{\Phi}\Phi 
+\left( \frac{1}{2} \int  d^2 \theta {\cal W} +\mbox{H.c}\right)
\label{one}
\eeq
where
\beq
{\cal W} = - iM^3\ln\Phi\,.
\label{two}
\eeq
The parameter $M$ can be always chosen to be real. The scalar 
potential $V$
is proportional to $|\phi|^{-2}$ (a ``mountain peak" centered at the 
origin 
in the space of fields). The run-away behavior is obvious.

For the wall-like solutions (i.e. static field configurations depending only 
on one coordinate $z$) the condition of the BPS saturation takes the 
form~\cite{DS1}
\beq
\frac{\partial\phi}{\partial z} = \frac{\partial\bar{\cal 
W}}{\partial\bar\phi}\,.
\label{three}
\eeq
It is quite obvious that, given the superpotential~(\ref{two}),
the solution of Eq.~(\ref{three}) of the winding-phase  type is
\beq
\phi_0 (z) = m e^{i\alpha(z)}\,,\qquad \alpha(z) =\frac{M^3}{m^2} z\,.
\eeq
For convenience we assumed $m$ to be real; its absolute value is 
arbitrary (one can always pass to a complex $m$ by a phase rotation 
of $\Phi$). The solution~(\ref{three}) preserves two out of four 
supercharges. The energy functional can be written as
\beq
E = \int d^3 x \left\{ \left| \frac{\partial\phi}{\partial z} 
+i\frac{M^3}{\bar\phi}\right|^2+ \left| \frac{\partial\phi}{\partial x} 
\right|^2+ \left| \frac{\partial\phi}{\partial y} \right|^2
+\left(\frac{\partial {\cal W}}{\partial z} 
+\mbox{H.c.}\right)\right\}\,.
\label{four}
\eeq
The corresponding vacuum energy density is
\beq
{\cal E} = 2\,\frac{M^6}{m^2}\, . 
\eeq 
Equation (\ref{four}) explicitly demonstrates that the system is 
stable 
under spatially localized perturbations. Indeed, if $\phi = \phi_0 
+\delta\phi$, and $\delta\phi$ vanishes at infinity,
\beq
\delta E= \int d^3 x \left\{ \left| \frac{\partial(\delta \phi )}{\partial 
z} 
-i\frac{M^3}{\bar\phi^2}\delta\bar\phi\right|^2+ \left| 
\frac{\partial(\delta \phi )}{\partial x} 
\right|^2+ \left| \frac{\partial(\delta \phi )}{\partial y} 
\right|^2\right\}\,. 
\eeq
There are no negative modes.
Thus, we get  a continuous family of vacua with a constant 
energy density labeled by the parameter $m$. 

 Another way to understand the stability is by compactifying
the coordinate $z$ on a circle of the radius $R$ and then taking the limit
$R \rightarrow \infty$. For finite $R$ only a discrete number of
solutions is allowed $M^3/m^2 = n/R~~(n=1,2..)$. Thus,
$M^3/m^2$ is a topologically conserved winding number density which
guaranties the stability of the configuration. Now, taking the
limit $R,n \rightarrow \infty$ with $n/R$ fixed we recover Eq. (4).

Now, let us discuss a model presenting a classic example of the 
spontaneous supersymmetry breaking (the O'Raifeartaigh 
mechanism~\cite{OR}). It includes three chiral superfields,
$\Phi_{1,2,3}$, with the superpotential
\beq
{\cal W} = \lambda_1\Phi_1(\Phi_3^2-M^2) +\mu \Phi_2\Phi_3\,.
\label{orsp}
\eeq
Again, it is convenient to choose the parameters $\lambda_1 , \,\, M$
and $\mu$ real. 
Superpotential of the type (\ref{orsp})  appear in the low-energy limit 
of various 
gauge field theories with matter.
If $M^2<\mu^2/(2\lambda_1^2)$ the minimum of energy is achieved 
at $\phi_2=\phi_3=0$ and $\phi_1$ undetermined. At the minimum 
the $F_1$ term does not vanish, $F_1=\lambda_1 M^2$, so that
supersymmetry is broken, and the vacuum energy density ${\cal E} =
\lambda_1^2 M^4$. Note that the flat direction along $\phi_1$
is lifted by the $Z$ factor arising as a (perturbative) quantum 
correction to the kinetic term. 

Thus, in  the flat vacuum SUSY is totally broken. Instead, one can try 
to 
find a BPS saturated wall-like solution preserving one half of SUSY. 
The 
BPS saturation conditions now take the form
\beq
\frac{\partial\phi_i}{\partial z} = \frac{\partial\bar{\cal 
W}}{\partial\bar\phi_i}\,, \qquad i=1,2,3\,.
\label{seven}
\eeq
They have an obvious solution 
\beq
\phi_1 = - \lambda_1 M^2 z\,,\quad \phi_2 = 0, \quad \phi_3 = 0\,. 
\label{eight}
\eeq
The vacuum energy density for this field configuration is 
\beq
{\cal E} = 2\,\lambda_1^2 M^4\, ,
\label{nine}
\eeq 
i.e. twice higher than in the Lorentz-invariant non-supersymmetric 
vacuum. Nevertheless, the configuration  (\ref{eight}) is absolutely 
stable 
under all localized deformations, much in the same way as in the 
case of the winding phase configuration of the previous example. 

In both cases the residual one half of SUSY guarantees that the 
fermion-boson degeneracy persists for the excitation modes in the 
given backgrounds. In the latter case, Eq. (\ref{eight}), the excitation 
modes from $\Phi_3$ are localized in the $z$ direction.

The {\em total} vacuum energy   gets no quantum corrections due to 
the BPS-saturated nature of the wall-like solutions considered.
However, the $z$-independence of the vacuum energy {\em density}
is lifted, generally speaking, by quantum corrections to the kinetic 
term. In the weak coupling regime these quantum corrections are small,
however. 

The mathematical foundation for the existence of the spatially 
delocalized vacuum configurations with the residual supersymmetry and 
a (classically) constant energy density, which we present here,  is the 
central extension of the ${\cal N} = 1$ superalgebra with an {\em 
infinite} 
central charge,
\beq
\{Q_\alpha Q_\beta\} =\Sigma_{\alpha\beta} {\cal Z}
\eeq
where $\Sigma_{\alpha\beta}$ is proportional to the area tensor
in the plane perpendicular to the $z$ direction, and 
\beq
{\cal Z} = 2\left\{ {\cal W}(z=L) - {\cal W}(z=-L) \right\} \, \propto
{\rm  const}\, L \to\infty 
\eeq
in both models considered. 
This is a natural generalization of the central extensions of the ${\cal 
N} = 1$ superalgebra with a finite value of the central charge found 
and discussed previously~\cite{DS2}. 
(Note that when we speak of the finite/infinite central charge we do
{\em not} include in ${\cal Z}$ a trivial area factor
$\Sigma_{\alpha\beta}\propto A$, which is, of course, infinite since the
wall area $A\to\infty$. For a recent discussion of a general  theory of
the tensorial central charges in various superalgebras in three and four
dimensions see~\cite{FP}.) In the examples discussed
in~\cite{DS2} the walls interpolate between a discrete set of
vacua related to each other by phase transformations.
Therefore, the central charge can take one of several possible
(finite) values from a discrete finite set. Whereas in the present case
there exists a symmetry of the model {\em per se}, or of the vacuum 
state, under which the superpotential
${\cal W}$ gets a  shift. This explains why the central charge is 
infinite. 

In the non-supersymmetric context stable soliton-like vacua in the
theories without the Lorentz-invariant vacua were discussed
in Ref.~\cite{Jackiw} and, more recently, in 
Ref.~\cite{CV}, where the question was raised as to the relevance of such 
configurations in the cosmological setting. In supersymmetric world
the models with no supersymmetric vacuum are abundant. The 
 vacua of the type we discuss here may play an important role
in the description of the cosmology emerging, in particular, 
 in the context of the TeV Planck scale scenario~\cite{DD}. 
First ideas in this direction will be presented in Ref.~\cite{DSnew}.

\vspace{0.5cm}

{\bf Acknowledgments}: \hspace{0.2cm}

This work was supported in part by DOE under the grant number
DE-FG02-94ER40823.

\end{document}